\documentclass[useAMS,usenatbib,usegraphicx]{mn2e}
\usepackage{amsmath}
\usepackage{amsfonts}
\usepackage{amssymb}
\usepackage{mathrsfs}

\title[Low frequency radio spectrum of LS\,5039 during periastron and apastron passages]
{Low frequency radio spectrum of LS\,5039 during periastron and apastron passages}

\author[Subir Bhattacharyya Sagar Godambe Nilay Bhatt Abhas Mitra Manojendu Choudhury]
{Subir Bhattacharyya$^1$\thanks{E-mail:subirb@barc.gov.in} Sagar Godambe$^1$   
 Nilay Bhatt$^1$  Abhas Mitra$^1$  Manojendu Choudhury$^2$
 \\
$^1$Astrophysical Sciences Division, Bhabha Atomic Research Centre, Mumbai 400085, India \\
$^2$Homi Bhabha Centre for Science Education, TIFR, Mumbai 400094, India 
}

\begin{document}


\pagerange{\pageref{firstpage}--\pageref{lastpage}} \pubyear{2008}

\maketitle

\label{firstpage}

\begin{abstract}

We have recently studied LS\,5039, a gamma-ray binary,  with
Giant Meterwave Radio Telescope (GMRT) during its periastron and apastron passage. 
The results presented here show that the spectra are inverted at the low frequency
and the flux densities do not differ significantly for two different orbital phases. 
Assuming that the free-free absorption of radio in stellar wind environment is responsible
for the optically thick radio emission we calculated the free-free absorption optical depth
and constrained the height of the radio emitting region from the orbital plane. The height 
is found to be around 1.6 AU for a spherical stellar wind geometry. This estimate may
change if the stellar wind is focussed or the radio absorption is due to synchrotron self-absorption.
\end{abstract}


\section{Introduction}
The major unresolved issues regarding the gamma-ray binary LS\,5039 are $(i)$
the nature of the compact object is not known, $(ii)$ till date there is no
convincing evidence in favour of the existence of any accretion disc in the 
source, and $(iii)$ the nature of the stellar wind is not clearly understood.

\noindent \cite{lph06} showed LS\,5039 to
be a high mass X-ray binary with a ON6.5V(f) type primary star of mass 22.9~M$_{\odot}$.
\cite{swain04} and \cite{cas05} studied
the source in UV and optical. Through UV and optical spectroscopy \cite{swain04}
 found P Cygni line profiles in the UV Nv $\lambda$ 1204 and CIV
$\lambda$ 1550 lines indicating the presence of strong wind. \cite{cas05} studied the optical
H Balmer and HeI and HeII lines and determined the orbital parameters. 

\noindent In a recent study carried out by \cite{sart11} using simultaneous 
observations with the space based Canadian Microvariability and Oscillation of Stars
({\emph MOST}) and ground based 2.3 m optical telescope Australian National University (ANU),
the stellar wind properties and the orbital parameters were studied extensively. Orbital 
parameters were better constrained as compared to \cite{cas05}. The eccentricity of the orbit
is found to be 0.24 as comapred to 0.35 by \cite{cas05} and 0.48 by \cite{swain04}.
Since the compact object mass was found to be $> 1.8M_{\odot}$, we find it is difficult to conclude about 
the nature of the compact object because with favourable equation of states the masses of neutron 
stars can easily approach such a value.

\noindent From their observations of H$\beta$ and HeI spectral lines \cite{sart11}
revealed that equivalent width (EW) varies with the orbital phase and, particularly, EW is lower at
inferior conjuction. This observation suggests that the stellar wind is possibly focussed towards 
the compact object instead of being spherically symmetric. A focussed wind is also observed in case 
of Cyg X-1 (\cite{gb86,mill05}), a high mass X-ray binary and 
it was modelled by \cite{gb86}. Even though the presence of focussed stellar wind in LS\,5039 needs further
study and is to be modelled properly, but it possibly has direct implications on the results we present in this 
paper.

\noindent In our previous work 
(\cite{god08}, Paper I), it was shown that the spectrum in the low frequency range is optically 
thick and there is a possible spectral turn over at around 1 GHz. The spectral index was found to be 
$-0.749\pm0.111$. The slope of the optically thick spectrum differs from the ideal condition of synchrotron 
self-absorbed spectrum which shows a spectral slope of $\frac{5}{2}$. The optically thick spectrum in
radio can also be generated by free-free absorption of radiation in presence of the stellar wind from
the hot and massive companion star. Considering the above two possible scenarios of absorption of radio 
spectrum \cite{br09}
 constrained the physical conditions of the radio emitting region in LS\,5039 jet. In case
of free-free absorption of radio wave in the stellar wind of the companion the magnetic field in the
emission region was estimated to be in the range $(3-30)\times10^{-3}$ G while the size and the location of the 
emitter will be $(3-4.5)\times10^{13}$ cm and $4.5\times10^{13}$ cm respectively. If SSA was responsible 
for the optically thick radio spectrum then the magnetic was found to be in the range $10^{-2}-1$ G. The size
of the emitting region would be $\sim (4.5-10) \times 10^{13}$ cm and location of the source would be at 
$\gtrsim (4.5-10)\times10^{13}$ cm.  Hence it is important to understand the actual process responsible for
the absorption of radio emission.

\noindent Here we present the results of radio observations of LS\,5039 taken with Giant Meterwave Radio 
Telescope in the 200 -- 1280\,MHz frequency range. We study the low frequency spectrum
during the periastron and apastron passage of the compact star and attempt to understand its possible 
consequences. The observation and the data analysis are discussed in the next section. The results are
discussed in Section 3 and finally we conclude the paper in Section 4. 
\begin{table}
\caption{Log of GMRT observations} \label{table1}
\label{Table:ObsLog}
\begin{center}
\begin{tabular}{|c|c|c|}
\hline
MJD	& Frequency   	     	& Phase \\
	& (MHz)         	&range  \\
\hline
54573   & 234           &0.63 -- 0.70\\
	& 605.2		&0.63 -- 0.70 \\
54670   & 1280          &0.65 -- 0.72 \\
54684	& 1280		&0.98 -- 0.05 \\
54688   & 605.2		&0.99 -- 0.06 \\
\hline
\end{tabular}
\end{center}
\end{table}
\begin{figure}
\begin{center}
\includegraphics[width=0.4\textwidth,height=0.35\textheight,angle=-90]{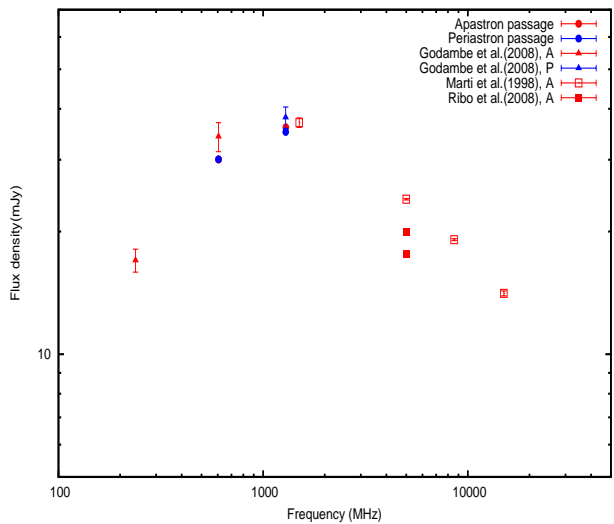}
\caption{Radio spectrum of LS\,5039 as observed with GMRT} \label{spec}
\end{center}
\end{figure}

\section{Observations and Data reduction}
\noindent The GMRT consist of 30 steerable antennas of 45\,m diameter in an approximate 
`Y' shape similar to the VLA but with each antenna in a fixed position. Fourteen 
antennas are randomly placed within a central 1 km $\times$ 1 km square (the 
`Central Square') and the remaining antennas form the irregular Y shape (six on
each arm) over a total extent of about 25 km. We refer the reader for details
about the GMRT array to http://gmrt.ncra.tifr.res.in and \cite{swarup91}. 
We observed LS\,5039 at 1280\,MHz on February 23, 2006 and simultaneously observed 
at 234 and 614\,MHz on March 13, 2006. 

The observations were made in standard fashion, with each source observation 
(30 min) interspersed with observations of the phase calibrator (4 min) . 
The primary flux density calibrator was either 3C\,48 and 
3C\,286, with all flux densities being on the scale of \cite{bars77}. Either of
these flux calibrator was observed for 20 min at the beginning and end of each observing
session. 

The data recorded 
with the GMRT was converted to FITS format and analyzed with the Astronomical Image 
Processing System (AIPS) using standard procedures. During the data reduction, 
we had to discard a fraction of the data affected by radio frequency interference 
and system malfunctions. After editing the data were calibrated and collapsed into
a fewer channels.
A self-calibration on the data was used to correct
for the phase related errors and improve the image quality. The observation dates, 
the observed frequencies and the corresponding spectral phases are given in Table \ref{table1}.

\section{Results and Discussion}
The spectra obtained for the two orbital phase ranges corresponding to the periastron and
apastron passage are shown in Fig. \ref{spec} along with the data points from our previous
observations (Paper I, \cite{god08}) and archival data points from \cite{mar98}
and \cite{ribo99}. The flux points obtained from \cite{ribo99} do not match with the
points obtained by \cite{mar98}. This is because the observation by \cite{ribo99}
with VLBI resolved the source and we have used the flux values reported for the core. The 
overall flux from the core and the wings is of the same order as reported by \cite{mar98}.
It is evident that the spectra remain optically thick and 
the flux values at frequencies 605 
and 1280\,MHz are almost equal during the periastron (phase range : 0.65 -- 0.72) and apastron
(phase range : 0.99 -- 0.06) passages. Because of poor quality of data at 234\,MHz, we could 
not produce source image for both the observing spells. 
Compared to the flux values of our previous observations, the
present flux values fall within the error bars of the previous observations. 
It is to be noted that the exposure time for our previous observations was less compared to
that of the present observations.  

\noindent The optically thick spectrum in radio frequencies can be produced due to the absorption of
radiation either by free-free absorption or by synchrotron-self absorption in the emitting region 
of the source. The free-free absorption of radio waves may occur in LS\,5039 due to the presence
of stellar wind fron the primary star. The relativistic jet where the radio emission takes place 
by synchrotron process, moves through an environment of such a stellar wind. For the time being, let us assume that the 
wind from the companion star is spherically symmetric. If so, the free-free absorption opacity is given
by (\cite{rl86})    
\begin{equation}
\alpha_{\nu}^{ff} = 0.018 T^{-\frac{3}{2}} Z^2 n^2 \nu^{-2} \bar g_{ff}
\end{equation}
where T is the temperature of the stellar wind, n is the number density of the stellar wind and $\nu$ is
\begin{figure}
\begin{center}
\includegraphics[width=0.4\textwidth,height=0.2\textheight]{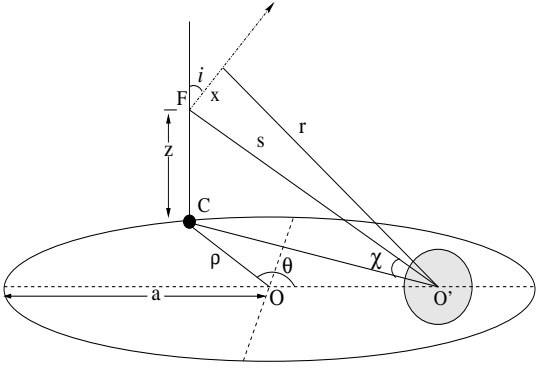}
\caption{Schematic diagram of binary orbit and the jet} \label{orbit}
\end{center}
\end{figure}the frequency of radiation. 

\noindent For a binary orbit inclined at an angle $i$ with respect to the sky plane, the jet will make an angle 
$i$ with the ditection of observation. It is assumed that the jet is perpendicular to the orbital plane. If the 
semi-major axis of the orbit is $a$ and the eccentricity is $e$ then the equation of the orbit is given by 
\begin{equation}\label{eq1} 
\rho^2 = \frac{a^2(1-e^2)}{(1-e^2 \cos^2\theta)}
\end{equation} 
where $\theta$ is the phase angle and $\theta\,=\,2\pi \phi$, $0\leq\phi\leq1$ is orbital phase. From Figure 
\ref{orbit}, using simple geometry, one can write
\begin{equation}\label{eq2}
O^{\prime}C^2 = \rho^2 + a^2e^2 - 2  \rho a e \cos\theta.
\end{equation}
If the radio emitting region is located at a height $z$ from the base of the jet then 
\begin{equation} \label{eq3}
s^2 = O^{\prime}C^2 + z^2 =  \rho^2 + a^2e^2 - 2  \rho a e \cos\theta + z^2
\end{equation}
In Figure \ref{orbit} $\angle O^{\prime}FC\, =\, (\frac{\pi}{2}-\chi)$, so
\begin{equation} \label{eq4}
r^2 = s^2 + x^2 - 2 s x \cos\left(\frac{\pi}{2}+(\chi-i)\right) 
\end{equation} 
Using equations (\ref{eq1}) and (\ref{eq3}) equation (\ref{eq4}) can be written as
\begin{eqnarray} \label{eq5}
r^2 = \frac{a^2(1-e^2)}{1-e^2 \cos^2\theta} + a^2e^2 - 2  \sqrt{\frac{a^2(1-e^2)}{1-e^2 \cos^2\theta}} a e \cos\theta + \nonumber \\
       z^2 + x^2 + 2 s x \sin(\chi-i)
\end{eqnarray}
Expressing all distances in the units of $a$, equation (\ref{eq5}) can be rewritten as
\begin{equation} \label{eq6}
{\bar r}^2 = {\bar K}^2(\theta) + {\bar z}^2 + {\bar x}^2 + 2 {\bar s} {\bar x} \sin(\chi-i)
\end{equation}
where ${\bar r}=\frac{r}{a}$, ${\bar z}=\frac{z}{a}$, ${\bar x}=\frac{x}{a}$ and 
\begin{equation} \label{eq7}
{\bar K}^2(\theta)= \frac{1-e^2}{1-e^2 \cos^2\theta} + e^2 - 2 e \sqrt{\frac{1-e^2}{1-e^2 \cos^2\theta}} \cos\theta   
\end{equation}
Using the geometry of Figure \ref{orbit} one can finally write 
\begin{equation} \label{eq8}
{\bar r}^2 = (\bar x + \mathscr{Z})^2 + \mathscr{B}^2
\end{equation}
where $\mathscr{Z} = \bar z \, \cos i \, - \, \bar K(\theta) \, \sin{i}$ and $\mathscr{B} = \bar z\, \sin{i} \,+\, \bar K(\theta) \, \cos{i}$.
For an aligned jet, where $i=0$, $\mathscr{Z} = \bar z$ and $\mathscr{B} = \bar K(\theta)$.

\noindent For a spherical wind the wind density at a distance $r$ from star is given by (\cite{puls96}),
\begin{equation}
n = \frac{\dot M}{4 \pi m_p r^2 v_{\infty} (1-\frac{R_{\star}}{r})} = \frac{\dot M}{4 \pi m_p a^2 {\bar r}^2 v_{\infty} (1-\frac{\bar R_{\star}}{\bar r})}.
\end{equation}   
Here $\dot M$ is the mass loss rate of the wind, $R_{\star}$ is the radius of the star, $v_{\infty}$ is the
terminal velocity of the wind at far away distance. The free-free absorption optical depth is given by
\footnote{Similar estimation of free-free optical depth is also done by \cite{zd11}.} 
\begin{equation}
\tau_{ff}(\bar z) = \frac{0.018 T^{-3/2} \nu^{-2} {\dot M}^2}{16 \pi^2 a^3 v_{\infty}^2 m_p^2}
\int_{0}^{\infty} {\bar r}^{-4} \left(1-\frac{\bar R_{\star}}{\bar r}\right)^{-2} d{\bar x}.
\end{equation}
As $r > R_{\star}$, so with the approximation that $(1-\frac{\bar R_{\star}}{\bar r})^{-2} \approx (1+\frac{2\bar R_{\star}}{\bar r})$, the optical
depth is given by 
\begin{align}
\tau_{ff}(\bar z)\; = \; {}&\frac{0.018 T^{-3/2} \nu^{-2} {\dot M}^2}{16 \pi^2 a^3 v_{\infty}^2 m_p^2} \, 
                    \Bigg\{
                    \frac{1}{2\,\mathscr{B}^3} \Bigg[\frac{\pi}{2} - \tan^{-1}\left(\frac{\mathscr{Z}}{\mathscr{B}}\right) - \nonumber \\ 
                    {}&\frac{1}{2}\,\sin\left(2\,\tan^{-1}\left(\frac{\mathscr{Z}}{\mathscr{B}}\right) \right) \Bigg] + 
                    \frac{2\,\bar R_{\star}}{\mathscr{B}^4} \Bigg[\frac{2}{3} - 
                    \sin\left(\tan^{-1}\left(\frac{\mathscr{Z}}{\mathscr{B}}\right) \right) \nonumber \\
                    +& \frac{1}{3} \, \sin^{3}\left(\tan^{-1}\left(\frac{\mathscr{Z}}{\mathscr{B}}\right) \right) \Bigg]  
                    \Bigg\}.
\end{align}   

\begin{figure}
\begin{center}
\includegraphics[width=0.4\textwidth,height=0.35\textheight,angle=-90]{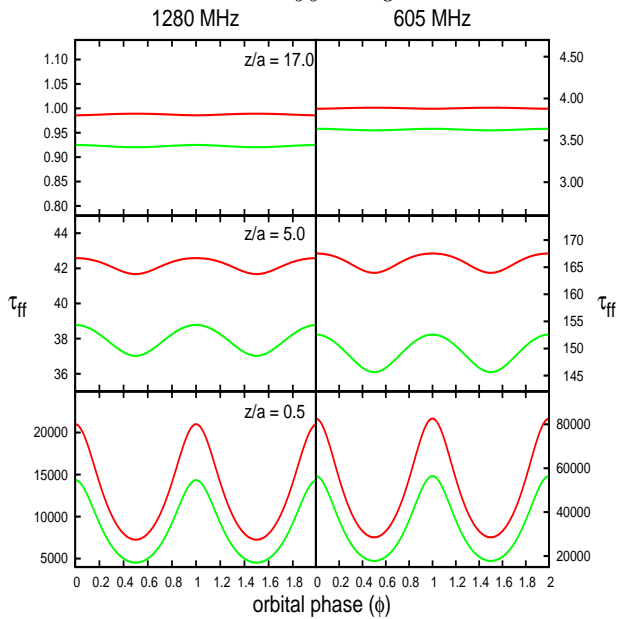}
\vspace{1.0cm}
\caption{Variation of free-free absorption optical depth with orbital phase for 605 MHz and 1280 MHz for three
different heights of the radio emitting region in the jet. \emph{Red curve}: $i=24^o$ and \emph{Green curve}
: $i=0^o$. Here the semi-major axis $a$ is taken as 20$R_{\odot}$}  \label{optdep1}
\end{center}
\end{figure}

\noindent For LS~5039, the parameter values are  $T\,=\,37500$K, $\dot M\,=\,10^{-7}\,M_{\odot}$ yr$^{-1}$, $a\,=\,20\,R_{\odot}$, $e\,=\,0.24$,
$R_{\star}\,=\,9.5\,R_{\odot}$, $i\,=\,24^o$ and $v_{\infty}\,=\,2440$ km s$^{-1}$ (\cite{swain04, cas05, swain11}).
The variation of optical depth with orbital phase for three different locations of the radio emitting
regions are shown in Figure \ref{optdep1} for frequencies 1280 and 605 MHz. For the inner jet it is evident that
the optical depth varies periodically. It is maximum during the periastron passage ($\phi\,=\,0$) and is minimum 
during apastron passage ($\phi\,=\,0.5$). This is true for both $0^o$ and $24^o$ inclination angle of the jet 
\emph{vis-a-vis} the inclination of orbital plane with respect to the sky plane. As the height of the radio
emission zone increases the periodic variation of the optical depth diminishes and essentially the optical depth 
becomes almost independent of the orbital phase for $\bar z \approx 17$ ($z\,=\,17a = 1.6$ AU). 
For $z \approx 1.6$ AU, the optical depth
for 1280 MHz, $\tau_{ff}(\nu=1280 \text{MHz}) \approx 1$ for the inclination angle 24$^o$. As the spectral turnover
is found to be around 1000 MHz, so $\tau_{ff} \approx 1$ is the necessary condition to constrain the emission height 
for the given physical condition.  Since the optical depth at this height becomes independent of the
orbital phase so the emitted radio flux densities will also be phase-independent. The approximately constant observed 
flux during periastron and apastron passages reported here does not contradict the present absorption model. 
For an aligned jet ($i=0$), the estimated height of the radio emission zone will be little less than
1.6 AU. These estimates indeed assume that the basic emission process remains unchanged for all orbital phases. It is
only the wind density changes for different orbital phases offering different optical depth to the emitted radiation.

\noindent However these estimations may change if $(i)$ the stellar wind is focussed or $(ii)$ the absortion is
due to synchrotron self-absorption in the jet. In fact recent observation of He I and H$\beta$ lines from 
LS\,5039 by \cite{sart11} indicates the presence of focussed wind in the binary system. \cite{sart11} 
observed significant changes in the H$\beta$ and He I spectral lines with the orbital phase. They attributed the 
low equivalent width of the spectral line to the focussing of the stellar wind towards the compact object due to its strong gravity.
\cite{br07} analysed the XMM-Newton data of LS\,5039 during the periastron and apastron passage
to study the effect of stellar wind on the X-ray absorption properties in this source. The spectral fitting of the
data gave a value of equivalent hydrogen column density ($N_H$) which is consistent with the interstellar value, for
all observation spells. \cite{br07} used the photoelectric absorption model \emph{phabs}. 
We reanalysed the same XMM-Newton 
data and fitted the spectrum with \emph{powerlaw} and the improved X-ray absorption model \emph{tbabs} using the spectral 
analysis tool, XSPEC. The X-ray absorption
cross-section used in the model \emph{tbabs} includes the cross-section for X-ray absorption by gas-phase ISM, by 
grain-phase ISM and by the molecules in the ISM. The fitted parameters for different observations are tabulated in
Table \ref{table2}. The value of $N_H$ obtained in our fitting is compatible to the ISM value. This again indicates
that the stellar wind does not contribute to the photoelectric absorption of the X-rays. This could be possible
if the wind geometry is not spherical. Thus if the wind is focussed
then the radio emission may take place at a lower height compared to the spherical wind scenario. So it is important to 
confirm the nature of stellar wind geometry for the companion star in LS~5039. 
In this context it is important to mention that Cyg X-1, a high mass X-ray binary, also exhibits focussed 
stellar wind from massive companion (\cite{gb86,mill05}).

\noindent Another possibility of absorption of the radiation is due to synchrotron self-absorption of radio emission
in jet. This scenario was 
discussed by \cite{br09} also. The absence of any orbital modulation of low frequency radio spectrum 
rule out this picture for the inverted radio spectrum of LS\,5039. The estimation by \cite{br09} assumed an ideal $\frac{5}{2}$ slope
in the self-absorbed region of the spectrum. But observation shows that the spectral slope of optically
thick portion of the spectrum deviates substatially from $\frac{5}{2}$ (Paper I). 
It requires the modeling of the spectrum possibly generated in an inhomogeneous jet which is beyond the 
scope of this paper.    
To estimate the physical parameters of the jet it is important to model the multiwavelength data from the 
source. Such a model, incorporating the data from the low-frequency radio to gamma-rays, will be used to self-consistently 
produce the inverted radio spectrum, to reproduce the spectral turn over at around 1 GHz and to fit the broadband 
spectra, will be discussed in our next work. 

\begin{table*}
\begin{minipage}{\textwidth}
\caption{Spectral parameters for XMM-Newton observations for two different absorption models} \label{table2}
\label{Table:ObsLog}
\begin{center}
\begin{tabular}{|c|c|c|c|c|c|c}
\hline
MJD	& ObsID       & Phase	     	& \multicolumn{2}{|c|}{\emph{tbabs*powerlaw}} & \multicolumn{2}{|c|}{\emph{wabs*powerlaw}} \\
\cline{4-7} \\
	&             &          	& photon index    & $N_H \times 10^{22}$ cm$^{-2}$  & photon index & $N_H \times 10^{22}$ cm$^{-2}$\\
\hline
54573   & 0151160201	& 0.53		& $1.50\pm0.03$   & $0.62\pm0.02$     	& $1.49\pm0.03$ 	& $0.63\pm0.03$ \\
54670   & 0151160301    & 0.54 		& $1.48\pm0.04$   & $0.66\pm0.03$ 	& $1.47\pm0.04$		& $0.67\pm0.03$ \\
54684	& 0202950201	& 0.51		& $1.53\pm0.03$	  & $0.69\pm0.02$ 	& $1.52\pm0.03$		& $0.71\pm0.02$ \\
54688   & 0202950301	& 0.03		& $1.63\pm0.04$   & $0.69\pm0.03$  	& $1.62\pm0.04$ 	& $0.71\pm0.03$ \\
\hline
\end{tabular}
\end{center}
\end{minipage}
\end{table*}

\section{Conclusion}

\noindent We observed the gamma-ray binary LS~5039 in the low frequency radio by using Giant Meterwave 
Radio Telescope (GMRT) during the periastron and apastron passage of the source. Observations were carried out 
at frequencies 234, 605 and 1280 MHz. The data at 234 MHz were discarded due to its poor quality. The analysed 
results show that the flux values at 605 and 1280 MHz are consistent with our previous observations and they do not
vary with the orbital phase. The spectrum is found to be inverted at low frequencies indicating an optically thick
radio spectrum. Present data, when combined with the archival data, indicate a spectral turn over around 1GHz (Figure 1). 
To understand the constancy of flux during apastron and periastron passages and to constrain the location of
radio emitting region we estimated the free-free absorption optical depth at radio frequencies due to the spherical 
stellar wind from the companion star in the binary system. From the free-free absorption optical depth estimates
it is argued that the height of the radio emitting region along the jet is around 1.6 AU.

\noindent It is to noted that the observed results presented here do not rule out the other possiblity of radio 
absorption by synchrotron self-absorption (SSA) process. If SSA is the responsible process then the above estimates may 
change. 

\noindent Also the estimation of  the height of the radio emitting region as described above assumes a spherical geometry
of the stellar wind. Recent observations indicate the presence of focussed wind in LS~5039 (\cite{sart11}). If a model
for focussed wind in this source is evolved and used to estimate the free-free absorption optical depth then also
the above estimates may change. Therefore to develop better understanding of radio emission and absorption
processes in LS~5039 it is important to have more simultaneous multiwavelength observations of the source.    

\section{Acknowledgement}
\noindent Authors acknowledge the reviewer Virginia McSwain for her constructive suggestions which improved the
the quality of the paper. The GMRT is a national facility operated by the National Centre for Radio Astrophysics of 
the Tata Institute of Fundamental Research. We acknowledge the help provided by the supporting 
staff of GMRT during the observations.

\bibliographystyle{mn2e}
\bibliography{ms2-rev}

\end{document}